\documentclass[12pt,thmsa]{article}
\usepackage{amssymb}

\usepackage{sw20jart}

\input{tcilatex}
\begin{document}

\title{Induced Spin-Currents in Alkali-Films}
\author{Gerd Bergmann, Funing Song and Doug Garrett \\
University of Southern California}
\date{\today }
\maketitle

\begin{abstract}
In sandwiches of FeK and FeCs the conduction electrons in the alkali metals
have a large mean free path. The experiments suggest that the specular
reflection for spin up and down electrons is different at the interface
yielding a spin current in the alkali film. The spin current is detected by
the anomalous Hall effect of Pb surface impurities.

 \newpage 
\end{abstract}

\section{Introduction}

Spin currents and the asymmetric scattering of electron spins by magnetic
moments and spin-orbit scatterers has attracted great scientific interest in
recent years. In the presence of a spin-orbit scatterer spin currents cause
an anomalous Hall effect \cite{B62}, \cite{H24}, \cite{B126}. The whole
complex of spin current, spin-orbit scattering and anomalous Hall effect
(AHE) is an area with many unsolved and very interesting problems.

In this paper we describe a number of experiments on quench-condensed
sandwiches of FeCs and FeK. Because of the low condensation temperature the
conduction electrons in the Fe have a very short mean free path of about 1 $%
nm$, but the electrons in the alkali films possess a mean free path $l_{K}$
which can be up to five time the film thickness. An example is shown in
Fig.1 where the mean free path of the K electrons is shown as a function of
the K thickness $d_{K}$. For a K thickness of 20 $nm$ the mean free path is
about $l_{K}=100$ $nm$, more than four times the thickness. This is rather
surprising since the Fe film is very disordered, and its mean free path is
only $l_{Fe}=1$ $nm$. Our conclusion is that the conduction electrons from
the K film barely enter the Fe film. The interface mostly reflects the
potassium electrons specularly. The detailed mechanism that decouples the
two films is not yet well understood. Because $l_{K}$ is so large we have to
assume that the potassium electrons are almost perfectly specularly
reflected at the interface. If the electrons would reach the disordered Fe
they would be scattered diffusively. This would yield according to Fuchs 
\cite{F31} and Sondheimer \cite{S36} an effective mean free path of the
order of the film thickness.

There is, however, a small transmission of electrons between the Fe and the
K films. In the presence of an electric field the electrons in the Fe do not
flow parallel to the electric field, but the current and electric field form
an angle due to the anomalous Hall effect (AHE). The current possesses a
component perpendicular to the electric field. If electrons cross the
interface between the Fe and the alkali film they transfer this
''anomalous'' current component into the non-magnetic alkali film and
introduce there an anomalous Hall conductance (AHC). Such an AHC has been
observed in sandwiches of FeCs and other sandwiches of a ferromagnet and an
alkali metal \cite{B121}.

Therefore sandwiches of Fe and Cs (or any other ferromagnet with any alkali
metal) represent a rather interesting system; (i) the electrons in the
alkali film behave almost as in a perfectly specularly reflecting film, (ii)
there is a weak transfer of electrons between the two films.

Since the chemical potentials for spin up and spin down electrons in the Fe
are different, one could expect a somewhat different reflection of the spin
up and down electrons at the interface. The bulk mean free path of the
conduction electrons in the alkali films is fairly large. Therefore the
effective mean free path (including surface scattering) is strongly
influenced by the degree of specular reflection at the interface. A
different specular reflectivity for spin up and down electrons would
therefore create a spin current in the alkali film. The question is: How
large is the spin current and how can it be detected?

It has been known for a long time that a spin current causes an AHE in the
presence of strong spin-orbit scattering. Ballentine and Huberman \cite{B62}
calculated the correction to the Hall constant in liquid Pb and Bi metals as
the consequence of the field-induced spin polarization and the spin-orbit
scattering by Pb or Bi atoms. Hirsch \cite{H24} introduced a spin Hall
effect due to spin-orbit scattering. One of the authors \cite{B126}
calculated exactly the AHE of a polarized electron gas in the presence of
spin-orbit scattering using Friedel phase shifts for the total angular
momentum.

The goal of this paper is to check whether sandwiches of FeCs possess a
detectable spin current in the Cs (in the presence of an electric field).
Our detection tool will be the spin-orbit scattering of Pb impurities which
we condense in situ on the free surface of the Cs. If the Cs possesses a
spin current close to its free surface then the Pb impurities will cause an
AHE. Therefore the anomalous Hall conductance (AHC) of the sandwich will
increase when the FeCs sandwich is covered with Pb.

For this investigation we quench condense at helium temperature a thin Fe
film with a thickness of about 10 $nm$ and a resistance of about 100 $\Omega 
$. After the Fe film has been annealed to 40 $K$ it is covered with a Cs
film. The thickness of the alkali film is varied between 5 $nm$ and 30 $nm$.
Finally the alkali film is covered with sub-mono-layers of Pb, starting with
0.01 atomic layers and then increasing the Pb coverage to 0.02, 0.05, 0.1,
0.2, 0.5 and 1 atomic layers. Each time the magneto-resistance and the Hall
resistance are measured in the field range between -7 and +7 $T$.

In Fig.2 the anomalous Hall resistance (after subtracting the linear Hall
resistance) is plotted for the FeCs sandwich and the same sandwich covered
with 2/100 atomic layers of Pb. The thicknesses of the Fe and Cs films are: $%
d_{Fe}=12.4$ $nm$, $d_{Cs}=29.6$ $nm$. One recognizes that the coverage with
0.02 atomic layers of Pb introduces a large AHC. We extrapolate the Hall
curves back to zero magnetic field. The Hall resistance of the alkali metals
is not perfectly linear in the magnetic field and the extrapolation has to
include this slight curvature. The extrapolated Hall resistance $R_{yx}^{0}$
is the anomalous Hall resistance for the artificial case of a sandwich in
field $B=0$ and the magnetization of the Fe polarized in the z-direction.

From the Hall resistance $R_{yx}^{0}$ and the resistance $R_{xx}$ (per
square) of the sandwich we calculate the Hall conductance $%
L_{xy}^{0}=R_{yx}^{0}/\left( R_{xx}\right) ^{2}$. The anomalous Hall
conductance is shown in Fig.3 as a function of the Pb coverage (in units of
atomic layers). It increases strongly with the first 1/100 of a mono-layer
of Pb. (Actually the increase in $R_{yx}^{0}$ is even stronger than in $%
L_{xy}^{0}$ since $R_{xx}$ increases as well with the Pb coverage \cite{B107}%
). Somewhere at a Pb coverage of about 0.03 atomic layers $L_{xy}^{0}$
reaches its maximum. For increasing Pb coverage the AHC shows a steep
decline with a minimum at about 0.3 atomic layers of Pb. Further increase in
the Pb coverage yields a soft increase of $L_{xy}^{0}$ which approaches a
saturation value. One appreciates the size of the AHC due to the Pb when one
compares it with the AHC of the underlying ferromagnetic Fe film. For the
disordered Fe film the AHC $L_{xy}^{0}$ has the values $6.6\times
10^{-5}\Omega ^{-1}$, for the FeCs sandwich $L_{xy}^{0}$ is $1.0\times
10^{-4}\Omega ^{-1}$ while the additional contribution due to $2/100$ atomic
layers of Pb is $5.0\times 10^{-4}\Omega ^{-1}$. Obviously we are dealing
here with a large effect.

In the next experimental step we want to make sure that it is the spin-orbit
scattering of Pb that produces the observed AHC. The Pb impurities are
strong spin-orbit scatterers because of their large nuclear charge.
Therefore we have performed additional experiments where similar FeCs
sandwiches are covered with sub-mono-layers of Ag and Mg. Ag and Mg
impurities possess a much smaller spin-orbit scattering than Pb impurities.
The results for the Ag coverage are included in Fig.3. In this case the AHC
does not show the pronounced peak. Instead the additional AHE is much
smaller and $L_{xy}^{0}$ approaches its saturation value on the scale of a
few hundredth of a mono-layer of Ag. The result for the Mg coverage is
similar. This confirms that the spin-orbit scattering of the Pb is essential
in the observation of the AHC.

In addition we have investigated the size of the AHC as a function of the Cs
thickness. The magnitude of the AHE due the Pb increases with increasing Cs
thickness. It also depends on the resistance of the Fe film. For Fe layers
with a thickness of 12 $nm$ and a resistance of $R_{xx}=36\Omega $ the $%
\Delta L_{xy}^{0}$ increases roughly linear with the Cs thickness. The shape
and in particular the width of the peak is independent of the Cs thickness,
the maximum is always at about 0.03 atomic layers of Pb.

It is difficult at the present time to describe the observed AHC
quantitatively since there are too many unknown parameters. But there are a
number of qualitative considerations that are useful.

We assume that an electron wave of spin up (down) is mainly specularly
reflected at the interface which decreases its amplitude by a factor $%
a_{\uparrow }^{i}$ ($a_{\downarrow }^{i}$). Both spin directions are
partially specularly reflected at the free surface with a change of
amplitude by $b^{s}$. Fuchs \cite{F31} and Sondheimer \cite{S36} derived an
expression for the conductivity of a thin film with the coefficient of
specular reflection $p$ at the surfaces. By extending this theory to a film
with specular reflection $p_{\uparrow ,\downarrow }^{i}=\left| a_{\uparrow
,\downarrow }^{i}\right| ^{2}$ on one surface (the interface) and $%
p^{s}=\left| b^{s}\right| ^{2}$ on the other surface one obtains for the
charge current conductivity and the spin current conductivity

\begin{eqnarray*}
\sigma _{c} &=&\sigma _{\uparrow }+\sigma _{\downarrow } \\
\sigma _{s} &=&\left( \sigma _{\uparrow }-\sigma _{\downarrow }\right)
\end{eqnarray*}
where $\sigma _{\uparrow ,\downarrow }=\left( ne^{2}/m\right) \left(
l_{\uparrow ,\downarrow }/v_{F}\right) $ are the conductivities for the two
spin directions. Here the effective mean free path $l_{\uparrow ,\downarrow
} $ =$l_{\uparrow ,\downarrow }\left( p_{\uparrow ,\downarrow
}^{i},p^{s},l_{0}\right) $ is a non-analytic function of the two
coefficients of specular reflection $p_{\uparrow ,\downarrow }^{i}$ and $%
p^{s}$ and the bulk mean free path $l_{0}$. (We define here the spin current
as the difference between the charge currents for spin up and spin down, $%
j_{s}=j_{\uparrow }-j_{\downarrow }$). As long as the spin current is not
changed by the condensation of the Pb impurities we expect a monotonic
increase of the AHC as a function of the Pb coverage which will be linear
for small coverages. In the experiment we see a maximum of $L_{xy}^{0}$ at
about 0.03 atomic layers of Pb. We can see two possible reasons why the spin
current is reduced for larger Pb coverages: \newline
$\left( \circ \right) $ \textbf{Spin rotation}: The spin-orbit scattering of
the Pb rotates the spin quantization direction. The electrons are no longer
in a spin up or down state but can have an arbitrary spin state $\left[
\alpha _{\uparrow }\left| \uparrow \right\rangle +\alpha _{\downarrow
}\left| \downarrow \right\rangle \right] $. Such an electron is specularly
reflected from the interface with an amplitude $\left[ a_{\uparrow }\alpha
_{\uparrow }\left| \uparrow \right\rangle +a_{\downarrow }\alpha
_{\downarrow }\left| \downarrow \right\rangle \right] $ . Its coefficient of
specular reflection is then $p^{i}=\left| a_{\uparrow }\alpha _{\uparrow
}\right| ^{2}+\left| a_{\downarrow }\alpha _{\downarrow }\right| ^{2}$. This
has two important consequences: (i) the spin current is reduced and (ii)
each interaction with the interface tries to restore the quantization in
z-direction. If $\left| a_{\uparrow }^{i}\right| ^{2}>\left| a_{\downarrow
}^{i}\right| ^{2}$ then the spin of the specularly reflected part of the
electron is rotated towards the up direction, while the diffusively
scattered component of the electron is rotated towards the down direction.

Our group has measured the spin-orbit scattering cross section of Pb on the
surface of Cs in the past \cite{B123}. We found a surprisingly large
spin-orbit scattering cross section of about $\sigma _{so}=0.5$*$\frac{4\pi 
}{k_{F}^{2}}\thickapprox 1.5\times 10^{-19}m^{2}$. This yields a spin-orbit
scattering time of about $10^{-12}s$ for 0.01 atomic layer of Pb on the
surface of Cs film with a thickness of 30 $nm$. This results in a spin-flip
rate is about $\frac{2}{3}*10^{12}s^{-1}$ which undermines the quatization
of the spin in z-direction and tends to reduce the spin current. The
spin-dependent reflection at the FeCs interface counteracts the
randomization of the spin. Both effects together yield a dynamic spin
distribution. \newline
($\circ $) \textbf{Diffuse surface scattering}: The condensation of (any)
impurities onto the Cs surface causes dramatic increase in the resistance of
the Cs film \cite{B107}. This reduces the coefficient of specular reflection 
$p^{s}$ on the free surface. In addition we have strong indications that the
(local) conductance close to the surface is strongly reduced. One possible
mechanism is that the surface impurities interact with the Cs atoms by
Friedel oscillations and cause disorder within 2 - 3 $nm$ of the surface.
This reduces strongly the charge and spin currents near the Pb atoms and
therefore the AHC would be strongly reduced.

We havew investigated which of the two mechanisms destroys the AHC at larger
Pb coverages by using instead of Pb another weaker spin-orbit scatterer. If
it is the spin-orbit scattering with its spin-flip processes that destroys
the spin current then we expect a smaller initial slope of the AHC as a
function of impurity coverage with a maximum at larger coverages.

We use Au impurities as an alternative spin-orbit scatterer. The nuclear
charge of Au is similar to that of Pb but since it is mainly an s-scatterer
its spin-orbit cross section is considerably smaller than that of Pb. We
cover an FeCs sandwich ($d_{Cs}=27.5$ $nm$) with sub-mono-layers of Au. In
Fig.4 its AHC $L_{xy}^{0}$ is plotted as a function of the Au coverage. The
sign of the AHC is reversed. This indicates that the asymmetry of the the Au
and Pb scattering have opposite signs. Obviously the AHE yields more
information about the spin-orbit scattering than the spin-orbit cross
section $\sigma _{so}$.

The Au impurities are less effective by a factor of four than the Pb
impurities. This confirms that the spin-orbit scattering of Au is smaller
than that of Pb. However, the extremum is for both impurities essentially at
the same coverage of about 0.03 atomic layers. This favors the explanation
that the impurity coverage reduces the coefficient of specular reflection at
the free surface and (or) that the impurities strongly reduce the local
conductance close to the surface.

In this paper we have covered sandwiches of FeCs with a few hundredth of a
mono-layer of Pb and observed a large anomalous Hall conductance. We can
explain the observed phenomena when we assume that the conduction electron
in the Cs have a different mean free path for spin up and down because their
reflection at the interface has different degrees of specularity. This
yields a spin current in the Cs and the strong spin-orbit scattering of the
Pb detects the spin current as an AHC. At the present time the Pb impurities
represent only a qualitative indicator of the size of the local spin
current. If the different phase shifts $\delta _{j,l}$ with $j=l\pm \frac{1}{%
2}$ were reliably calculated for Pb impurities on (in) Cs one could
quantitatively determine the size of the local spin-current at the Cs
surface using the results of ref. \cite{B126}. This system has the potential
for further experiments with spin currents in a non-magnetic metal. The
basic idea of sandwiching a ferromagnetic and a non-magnetic metal separated
by a barrier may be used to prepare other interesting systems.

Acknowledgment: The research was supported by NSF Grant No. DMR-0124422.

\newpage

\section{Figures}

\FRAME{dtbpF}{3.8795in}{3.1419in}{0pt}{}{}{f_1arb79.wmf}{\special{language
"Scientific Word";type "GRAPHIC";maintain-aspect-ratio TRUE;display
"USEDEF";valid_file "F";width 3.8795in;height 3.1419in;depth
0pt;original-width 1772.6875pt;original-height 1433.9375pt;cropleft
"0";croptop "1";cropright "1";cropbottom "0";filename
'C:/0aa/Aa_tex/Support/S_Arb79/F_1Arb79.WMF';file-properties "XNPEU";}}%
Fig.1: The mean free path of a K film quench-condensed on Fe as a function
of the K thickness.\FRAME{dtbpF}{4.0049in}{3.3356in}{0pt}{}{}{f_2arb79.wmf}{%
\special{language "Scientific Word";type "GRAPHIC";maintain-aspect-ratio
TRUE;display "USEDEF";valid_file "F";width 4.0049in;height 3.3356in;depth
0pt;original-width 1741.5625pt;original-height 1433.9375pt;cropleft
"0";croptop "1";cropright "1";cropbottom "0";filename
'C:/0aa/Aa_tex/Support/S_Arb79/F_2Arb79.WMF';file-properties "XNPEU";}}%
Fig.2: The anomalous Hall resistance of the FeCs sandwich before and after
coverage with 0.02 atomic layers of Pb. (The normal linear Hall resistance
is subtracted).

\FRAME{dtbpF}{3.4662in}{2.9663in}{0pt}{}{}{f_3arb79.wmf}{\special{language
"Scientific Word";type "GRAPHIC";maintain-aspect-ratio TRUE;display
"USEDEF";valid_file "F";width 3.4662in;height 2.9663in;depth
0pt;original-width 1677.1875pt;original-height 1433.9375pt;cropleft
"0";croptop "1";cropright "1";cropbottom "0";filename
'C:/0aa/Aa_tex/Support/S_Arb79/F_3Arb79.WMF';file-properties "XNPEU";}}%
Fig.3: The anomalous Hall conductance $L_{xy}^{0}$ of an FeCsPb and an
FeCsAg sandwich as a function of the impurity coverage (Pb or Ag given in
atomic layers).

\FRAME{dtbpF}{3.5146in}{2.8236in}{0pt}{}{}{f_5arb79.wmf}{\special{language
"Scientific Word";type "GRAPHIC";maintain-aspect-ratio TRUE;display
"USEDEF";valid_file "F";width 3.5146in;height 2.8236in;depth
0pt;original-width 1742.25pt;original-height 1326.4375pt;cropleft
"0.038772";croptop "0.950425";cropright "0.946061";cropbottom
"0.043075";filename
'C:/0aa/Aa_tex/Support/S_Arb79/F_5Arb79.WMF';file-properties "XNPEU";}}%
Fig.4: The additional anomalous Hall conductance $L_{xy}^{0}$ of an FeCsAu
sandwich as a function of the Au coverage.

\newpage

\section{Rest}

Spin-orbit impurities scatter conduction electrons with a left-right
asymmetry in the scattering amplitude. This asymmetry has the opposite sign
for opposite spins. As long as both spin orientation contribute equally to
the current they do not yield an anomalous ''Charge'' Hall effect. (However,
they yield opposite gradiants in the chemical potential for the two spin
orientations).

The Fe layer has, of course, a spin polarization and spin current. They
cause the AHC of the original Fe film. However, the spin polarization of the
Fe should not extend deeply into the Cs film, even in the presence of a
current. Furthermore one would expect that any polarization at the free Cs
surface by the Fe would decrease with increasing Cs thickness.

\newpage

\section{Theory}

The modeling of the observed AHE has to overcome a number of obstacles

\begin{itemize}
\item  The degree of specular reflection for spin up and down on the FeCs
interface and at Cs surface are not known.

\item  For the calculation of the conductance one has to use the
Fuchs-Sondheimer theory which yields non-analytic expression for the
conductance.

\item  Only for zero (or small) concentration of the spin-orbit scatterer Pb
the spin of the conduction electrons in the Cs is a good quantum number.

\item  The structural changes of the Cs surface du to the Pb impurities is
not well known and understood.
\end{itemize}

We strongly simplify the problem in the following way. Since a perfect
reflection at the interface or surface of the Cs does not reduce the
conductance we replace a slightly imperfect reflection by a scattering
lifetime. If $\delta _{\uparrow }^{i}$ is the fraction of spin up electrons
which is not specularly reflected at the interface then $\tau _{\uparrow
}^{i}=\delta _{\uparrow }^{i}\frac{2d_{cs}}{\overline{v_{Cs}}}$ where $%
d_{cs} $is the thickness of the Cs film and $\overline{v_{Cs}}$ is the
average of the electron velocity perpendicular to the surface. Similar we
define scattering lifetime at the surface (where both spins are equally
scattered) $\tau ^{s}=\delta ^{s}\frac{2d_{cs}}{\overline{v_{Cs}}}$. The
scattering lifetime in the interior of the Cs films is $\tau _{0}$. Then the
liefe time for spin up electrons is 
\begin{eqnarray*}
\frac{1}{\tau _{\uparrow }} &=&\frac{1}{\tau _{0}}+\delta _{\uparrow }^{i}%
\frac{\overline{v}}{d_{Cs}}+\delta ^{s}\frac{\overline{v}}{d_{Cs}} \\
\tau _{\uparrow } &=&\frac{d_{Cs}/\overline{v}}{\frac{d_{Cs}}{\tau _{0}%
\overline{v}}+\delta _{\uparrow }^{i}+\delta ^{s}}
\end{eqnarray*}
\[
G=\frac{ne^{2}}{m}d_{Cs}\left( \tau _{\uparrow }+\tau _{\downarrow }\right) 
\]
and the spin current density in the presence of an electric field $E$%
\[
j_{spin}=\frac{ne}{m}\left( \tau _{\uparrow }+\tau _{\downarrow }\right) E 
\]
\[
\frac{1}{1+x}-\frac{1}{2+x} 
\]
\[
\FRAME{itbpF}{3in}{2.0003in}{0in}{}{}{}{\special{language "Scientific
Word";type "MAPLEPLOT";width 3in;height 2.0003in;depth 0in;display
"USEDEF";plot_snapshots TRUE;function
\TEXUX{$\frac{1}{1+x}-\frac{1}{2+x}$};linecolor "black";linestyle
1;linethickness 1;pointstyle "point";xmin "0";xmax "5.07288";xviewmin
"0";xviewmax "5";yviewmin "0";yviewmax "1";viewset"XY";rangeset"X";recompute
TRUE;phi 45;theta 45;plottype 4;numpoints 49;axesstyle "normal";xis
\TEXUX{x};var1name \TEXUX{$x$};valid_file "T";tempfilename
'C:/0aa/Aa_tex/A_paper/Arb__/HS9SZP00.wmf';}} 
\]

Our group has measured the spin-orbit scattering cross section of Pb on the
surface of Cs in the past \cite{B123}. We found a surprising large
spin-orbit scattering cross section of about $\sigma _{so}=0.5$*$\frac{4\pi 
}{k_{F}^{2}}\thickapprox 1.5\times 10^{-19}m^{2}$. This yields a spin-orbit
scattering time of about $10^{-12}s$ for 0.01 atomic layer of Pb on the
surface of Cs film with a thickness of 30 $nm$. This results in a spin-flip
rate is about $\frac{2}{3}*10^{12}s^{-1}$ which undermines the quatization
of the spin in z-direction and tends to reduce the spin current. The spin
dependent reflection at the FeCs interface counteracts the randomization of
the spin. If for example the specular reflection of the spin up electrons is
larger than the for the spin down electrons then each interface reflection
rotates the spin of the specular reflected electrons towards the up
direction and the spin of the diffusely scattered part into the down
direction. 
\[
\]
We assume that the interface with the Fe reduces the amplitude in the
reflection by the factor $e_{+}$ and $e_{-}$ for spin up and spin down. Now
we consider an electron wave $\psi _{i}=\left| \mathbf{k}\right\rangle \chi $
with the spin function 
\[
\chi =\alpha \left| \uparrow \right\rangle +\beta \left| \downarrow
\right\rangle 
\]
After the reflection one obtains 
\[
\psi _{o}=e_{+}\alpha \left| \uparrow \right\rangle +e_{-}\beta \left|
\downarrow \right\rangle 
\]
Therefore the specular part of the reflection is $\left[ \left| e_{+}\alpha
\right| ^{2}+\left| e_{-}\beta \right| ^{2}\right] $ and the spin of the
wave changed from $\left[ \left| \alpha \right| ^{2}-\left| \beta \right|
^{2}\right] $ to $\left[ \left( \left| e_{+}\alpha \right| ^{2}-\left|
e_{-}\beta \right| ^{2}\right) /\left( \left| e_{+}\alpha \right|
^{2}+\left| e_{-}\beta \right| ^{2}\right) \right] $. If we set $e_{\pm
}=e\left( 1\pm \Delta \right) $ then we obtain for the probability 
\begin{eqnarray*}
&&\left[ \left| e_{+}\alpha \right| ^{2}+\left| e_{-}\beta \right|
^{2}\right] \\
&=&\left( e\left( 1+\Delta \right) \alpha \right) ^{2}+\left( e\left(
1-\Delta \right) \beta \right) ^{2} \\
&=&\allowbreak e^{2}+2e^{2}\left( \alpha ^{2}-\beta ^{2}\right) \Delta
+e^{2}\Delta ^{2}
\end{eqnarray*}
and for the final spin\thinspace 
\begin{eqnarray*}
&&\left[ \left( \left( e\left( 1+\Delta \right) \alpha \right) ^{2}-\left(
e\left( 1-\Delta \right) \beta \right) ^{2}\right) /\left( \left( e\left(
1+\Delta \right) \alpha \right) ^{2}+\left( e\left( 1-\Delta \right) \beta
\right) ^{2}\right) \right] \\
&=&\left( \alpha ^{2}-\beta ^{2}\right) +8\alpha ^{2}\beta ^{2}\Delta
-16\alpha ^{2}\beta ^{2}\left( \alpha ^{2}-\beta ^{2}\right) \Delta ^{2} \\
&=&\left( \alpha ^{2}-\beta ^{2}\right) +8\alpha ^{2}\beta ^{2}\Delta \left(
1-2\Delta \left( \alpha ^{2}-\beta ^{2}\right) \right)
\end{eqnarray*}
We use the linear part.

The electron with the opposite spin 
\[
\chi =\beta \left| \uparrow \right\rangle -\alpha \left| \downarrow
\right\rangle 
\]
is transferred into 
\begin{eqnarray*}
&=&e_{+}\beta \left| \uparrow \right\rangle -e_{-}\alpha \left| \downarrow
\right\rangle \\
&=&e\left[ e_{+}\beta \left| \uparrow \right\rangle -e_{-}\alpha \left|
\downarrow \right\rangle \right]
\end{eqnarray*}
with the probability 
\begin{eqnarray*}
&&\left[ \left| e_{+}\beta \right| ^{2}+\left| e_{-}\alpha \right|
^{2}\right] \\
&=&\left( e\left( 1+\Delta \right) \beta \right) ^{2}+\left( e\left(
1-\Delta \right) \alpha \right) ^{2} \\
&=&\allowbreak e^{2}-2e^{2}\left( \alpha ^{2}-\beta ^{2}\right) \Delta
+e^{2}\Delta ^{2}
\end{eqnarray*}
and the expectation value of $s_{z}$ equal to 
\begin{eqnarray*}
&&\left[ \left( \left| e_{+}\beta \right| ^{2}-\left| e_{-}\alpha \right|
^{2}\right) /\left( \left| e_{+}\beta \right| ^{2}+\left| e_{-}\alpha
\right| ^{2}\right) \right] \\
&=&\left( \left( e\left( 1+\Delta \right) \beta \right) ^{2}-\left( e\left(
1-\Delta \right) \alpha \right) ^{2}\right) /\left( \left( e\left( 1+\Delta
\right) \beta \right) ^{2}+\left( e\left( 1-\Delta \right) \alpha \right)
^{2}\right) \\
&=&-\left( \alpha ^{2}-\beta ^{2}\right) +8\alpha ^{2}\beta ^{2}\Delta
+16\alpha ^{2}\beta ^{2}\left( \alpha ^{2}-\beta ^{2}\right) \Delta ^{2}
\end{eqnarray*}

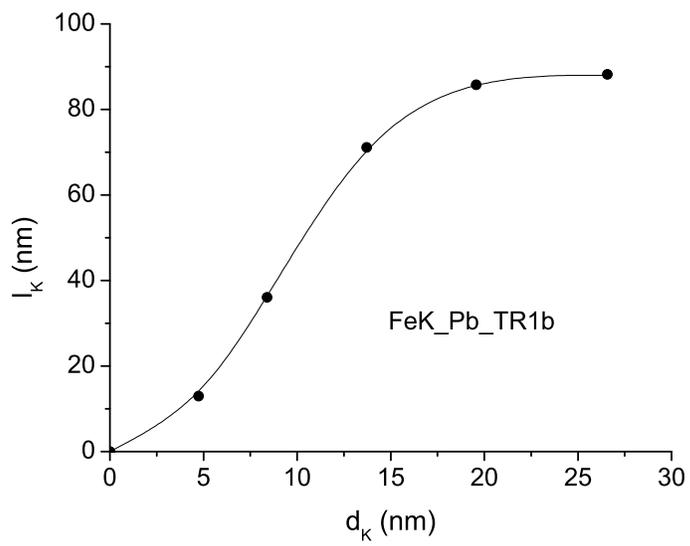
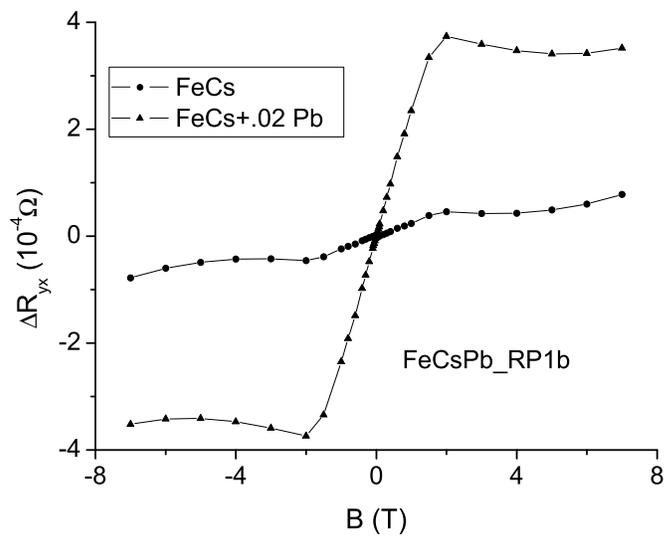
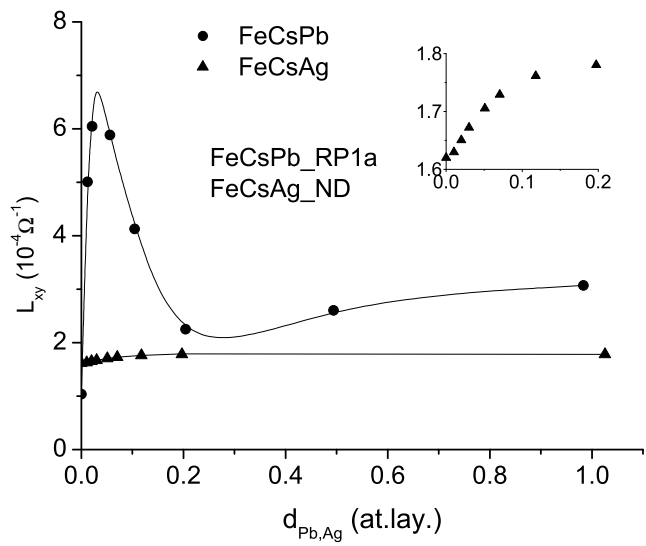
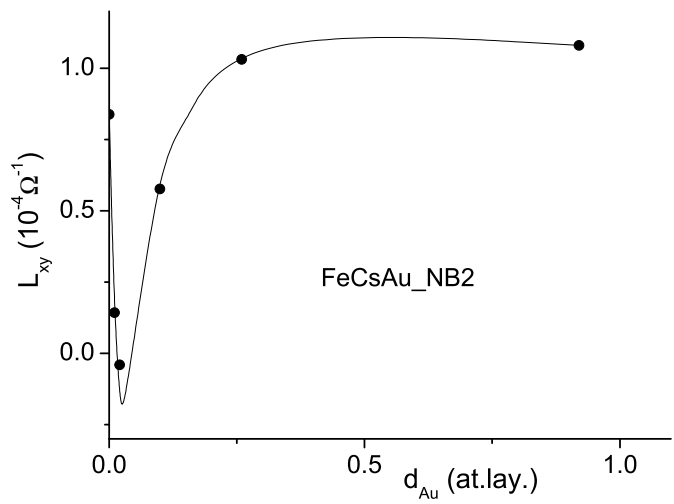
\end{document}